\documentclass[aps, prapplied, showpacs, reprint, preprintnumbers,amsmath,amssymb,superscriptaddress,floatfix]{revtex4-2}
\usepackage[]{graphicx}      
\usepackage{bm}             	
\usepackage{times}			
\usepackage{tabularx}
\usepackage{hyperref}
\usepackage{color}
\usepackage{dcolumn}		
\usepackage{amsmath} \usepackage{amssymb} \usepackage{amsfonts} \usepackage{times}
\usepackage{tabularx} \usepackage{xcolor,colortbl} \usepackage{amsmath}
\usepackage{tikz,xcolor,hyperref}
\definecolor{lime}{HTML}{A6CE39}
\DeclareRobustCommand{\orcidicon}{
	\begin{tikzpicture}
	\draw[lime, fill=lime] (0,0) 
	circle [radius=0.16] 
	node[white] {{\fontfamily{qag}\selectfont \tiny ID}};
	\draw[white, fill=white] (-0.0625,0.095) 
	circle [radius=0.007];
	\end{tikzpicture}
	\hspace{-2mm}
}

\foreach \x in {A, ..., Z}{\expandafter\xdef\csname orcid\x\endcsname{\noexpand\href{https://orcid.org/\csname orcidauthor\x\endcsname}
			{\noexpand\orcidicon}}
}

\makeatletter
\newcommand\xleftrightarrow[2][]{%
  \ext@arrow 9999{\longleftrightarrowfill@}{#1}{#2}}
\newcommand\longleftrightarrowfill@{%
  \arrowfill@\leftarrow\relbar\rightarrow}
\makeatother

\begin{document}

\title{Spin waves involved in three-magnon splitting in synthetic antiferromagnets}
\author{Asma Mouhoub 
}
\author{Nathalie Bardou}
\author{Jean-Paul Adam}
\affiliation{Universit\'e Paris-Saclay, CNRS, Centre de Nanosciences et de Nanotechnologies, Palaiseau, France}
\author{Aur\'elie Solignac} 
\affiliation{SPEC, CEA, CNRS, Universit\'e Paris-Saclay, 91191 Gif-sur-Yvette, France}
\author{Thibaut Devolder}
\email{thibaut.devolder@cnrs.fr}
\affiliation{Universit\'e Paris-Saclay, CNRS, Centre de Nanosciences et de Nanotechnologies, Palaiseau, France}
\date{\today}

\begin{abstract}
An important nonlinear effect in magnonics is the 3-magnon splitting where a high frequency magnon splits into two magnons of lower frequencies. Here, we study the 3-magnon splitting in spin wave conduits made from synthetic antiferromagnets. By combining inductive excitation, inductive detection, Brillouin Light Scattering imaging of the spin waves and analytical modeling based on conservation laws, we elucidate the nature of the spin waves involved in this process. We show in particular that low order optical spin waves propagating along the conduit can split in doublets of non-degenerate acoustic spin waves that have a standing wave character in the confined direction and unsymmetrical wavevectors in the direction of the spin wave conduit. Generally, several splitting channels run in parallel. The rules governing the three-magnon splitting and its interplay with the mode confinement have consequences for the applications in non-linear microwave signal processing based on spin waves.
\end{abstract}

\maketitle

Magnetization dynamics is intrinsically non-linear. Even the simplest dynamical state --the uniform ferromagnetic resonance (FMR)--  was recognized to be unstable at high signal powers \cite{suhl_theory_1957} soon after its discovery. The magnons are indeed the eigenmodes of the linearized dynamics of magnetic bodies: when substantially populated, magnons interact nonlinearly through scattering processes \cite{gurevich_magnetization_1996}. The three-magnon splitting (3MS) is the lowest order magnon-magnon interaction, where an initial high-frequency magnon splits into two new ones of lower frequencies \cite{boardman_three-_1988}. So far 3MS was measured mostly in extended films \cite{mathieu_brillouin_2003} where a magnetostatic surface wave typically splits into two backward volume SWs \cite{schultheiss_direct_2009, liu_time-resolved_2019}. This proceeds through a well-understood process,  most often degenerated \cite{ordonez-romero_three-magnon_2009} with applications in microwave signal processing \cite{harris_modern_2012}. 

The past decade has seen the development of 3MS studies in patterned structures, with a focus on dots in the vortex state \cite{schultheiss_excitation_2019}. There, selection rules lead to well-defined 3MS channels among the eigenmodes \cite{verba_theory_2021} which can be harnessed for unconventional computing \cite{korber_pattern_2023}. We can anticipate that 3MS in more general confined systems is also subject to stringent rules of the involved frequencies, with potential applications in non-linear signal processing. Unfortunately, the technical difficulty of generating and measuring SWs in multilayers and nanostructures formed from multilayers \cite{sheng_nonlocal_2023} is such that little is known about the 3MS in these systems.

Synthetic antiferromagnets (SAFs) are particularly adequate for studies on 3MS \cite{sud_electrically_2025}. Indeed SAFs possess two families of SWs that are the propagating counterparts of the acoustic (ac) and optical (op) modes of a two-flat-spin system \cite{stamps_spin_1994}. We refer to these modes as the acoustic and optical SWs. Their frequencies can be tuned by the field to satisfy energy conservation in 3MS and parallel pumping can be used to populate selectively the sole optical family, such that every detected SW from the acoustic family can be traced back to a 3MS event. In addition, the acoustic family of SWs in SAF can possess unidirectionality \cite{thiancourt_unidirectional_2024}: every acoustic SW generated by 3MS, no matter its wavevector, shall radiate energy toward the same half space. A single antenna can thus conveniently detect all the SWs generated by 3MS. 

Our aim is to determine the SWs involved in 3MS in SAFs, when an optical magnon pumped by an rf field at a frequency $\omega_\textrm{pump}=\omega_\textrm{op}$ splits into two acoustic magnons of frequencies $\omega_\textrm{ac,1}$ and $\omega_\textrm{ac,2}$. The measurement are done both inductively and by Brillouin Light Scattering microscopy ($\mu$BLS). We then describe the 3MS phenomenology: the splitting is most often non-degenerate (i.e. $\omega_\textrm{ac,1} \neq \omega_\textrm{ac,2}$) and can proceed through several channels, with different quantization signatures. A model based on conservation laws reproduces the experimental trends. The magnons generated by 3MS have a standing wave character with most often effective wavevectors that are neither collinear to each others nor to the initial optical magnon. 
\begin{figure*} \includegraphics[width=14 cm]{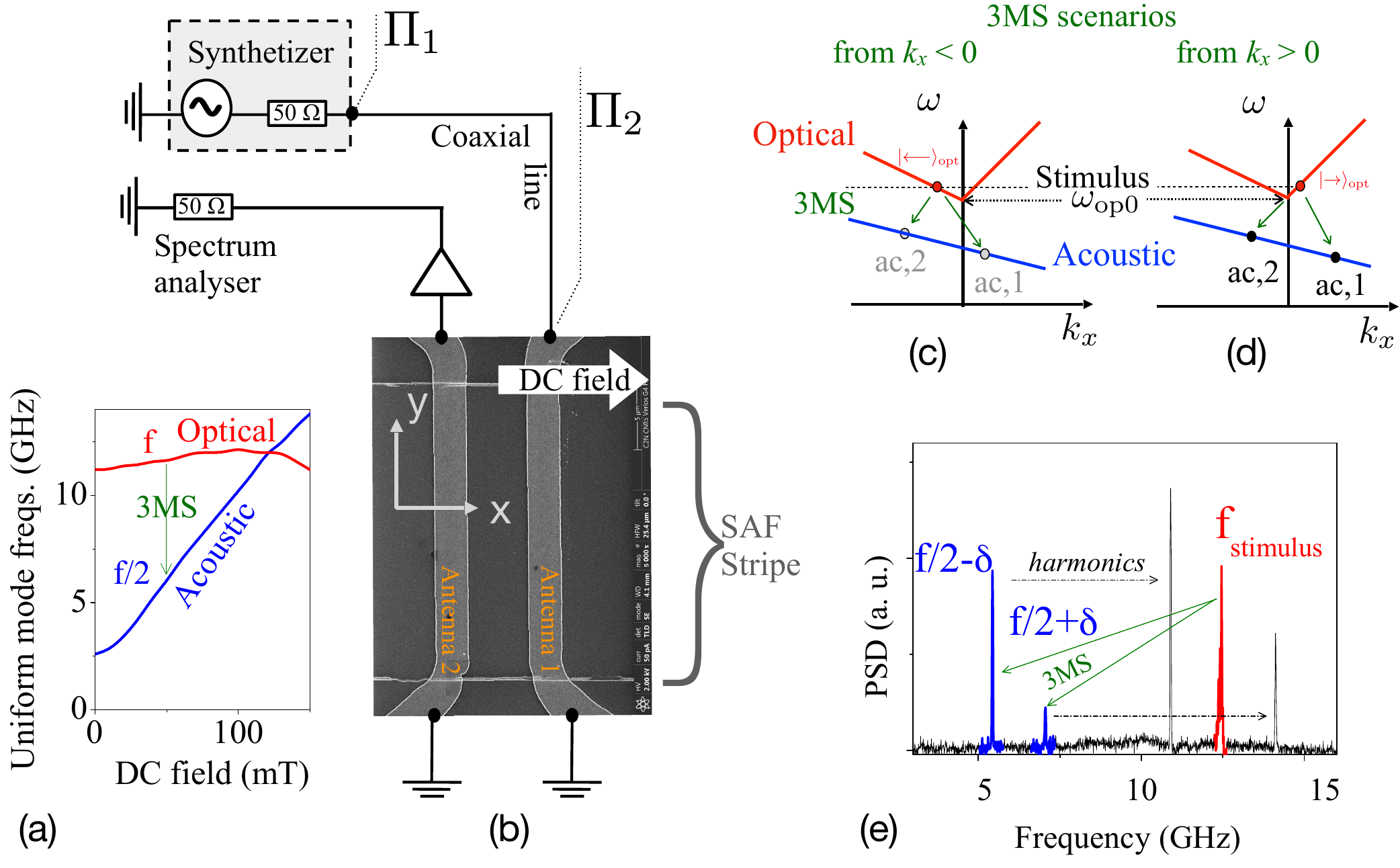}
\caption{Overview of the methods. (a):  
VNA-FMR measurement of the frequencies of the uniform acoustic and optical modes on an unpatterned SAF. The green arrow sketches the field required to get $\omega_\textrm{op}=2 \omega_\textrm{ac}$ at $\vec k=\vec 0 $. (b) Set-up and image of a sample with a stripe of width $w_\textrm{stripe}=20~\mu$m. The microwave losses $S_{21}(\omega)$ between the planes $\Pi_{1}$ and $\Pi_{2}$ were accounted for. (c, d): Illustration of the three-magnon scattering (3MS) processes in SAFs: the stimulus of frequency $f_\textrm{stimulus}$ excites a spin wave from the optical branch with either $k_x^\textrm{opt} <0 $ (panel c) or $k_x^\textrm{opt} >0 $ (panel d), which splits into two spin waves of the acoustic branch, with frequencies close to $f/2$. (e) Typical signature of this process as recorded with a spectrum analyzer at the lowest power where 3MS is observed. The black peaks are the harmonics of the blue ones.}
\label{Fig_1_Setup_Spectrum_qualitative_3MS} \end{figure*}

We use SAFs of composition Co$_{40}$Fe$_{40}$B$_{20}$ (17~nm) /Ru (0.7~nm) / Co$_{40}$Fe$_{40}$B$_{20}$ (17~nm). The properties includes a magnetization $M_s=1.35$ MA/m, a damping $\alpha=0.011$ and an interlayer coupling field of $\mu_0 H_j$= 100 mT \cite{seeger_inducing_2023, thiancourt_unidirectional_2024}. 
The SAF films are patterned into stripes of widths $w_\textrm{mag}$ from 2 to 20 $\mu \textrm{m}$ (Fig.~\ref{Fig_1_Setup_Spectrum_qualitative_3MS}).  The SAF are set in the scissors state by a static field $\mu_0 H_{x}$ adjusted to ensure that the uniform acoustic (ac) and optical (op) eigenmodes have quadratically related frequencies, i.e. that $\omega_\textrm{op0}=2 \omega_\textrm{ac0}$ [Fig.~\ref{Fig_1_Setup_Spectrum_qualitative_3MS}(a)]. 
The SAF stripes are finally covered by two straight antennas \footnote{The impedance $Z_\textrm{ant}$ of the widest antenna (width $w_\textrm{ant}=1.8~\mu \textrm{m}$) comprises a resistance of $9.2~\Omega$ and an inductance of 290 nH} that couple with SWs of wavevectors up to $k_\textrm{max}^\textrm{antenna} \approx 3 ~\textrm{rad}/\mu\textrm{m}$ in the $x$ direction. 
Fig.~\ref{Fig_1_Setup_Spectrum_qualitative_3MS}(b) describes our configuration. A microwave source feeds a first antenna to pump inductively optical magnons of frequency $\omega_\textrm{op}=\omega_\textrm{pump}$. 
The second antenna senses the SWs that travel below by feeding a spectrum analyzer configured to display the power spectrum that comes in excess compared to a reference situation chosen for its absence of non-linear processes. \\

When targeting at non-linear interactions the stimulus amplitude must be precisely determined. We thus measure the insertion losses $S_{21}(\omega)$ between the microwave source (reference plane $\Pi_1$, Fig.~\ref{Fig_1_Setup_Spectrum_qualitative_3MS}) of power $P_\textrm{source}$ and an impedance-matched termination placed instead of the first antenna (plane $\Pi_2$). The antenna peak current depends on $P_\textrm{source}$, on $S_{21}$, on the antenna impedance $Z_\textrm{ant}(\omega)$ and the characteristic impedance $Z_c=50~\Omega$. It is:
\begin{equation}
    I_\textrm{peak}(A)  =2 \sqrt2   \frac{\sqrt {Z_c}}{Z_\textrm{ant}+ Z_c}  {10^{\frac{P_\textrm{source} (\textrm{dBm}) - S_{21}(\textrm{dB}) } {20}- \frac32 }}
    \end{equation}  
When the power reaching the antenna is $P_\textrm{source} - S_{21}=0$ dBm at 11 GHz,  this leads to an antenna current of $I_\textrm{peak}=8.8$ mA.  Using Eqs.~A1-A4 of ref.~\onlinecite{devolder_propagating-spin-wave_2023}, this power leads to an rf field peaking at $\mu_0 h_x^\textrm{peak}=3.06$~mT at the SAF surface under the middle of the antenna. We will see that the thresholds of 3MS are of the order of $\mu_0 h_x^\textrm{peak}=1$~mT.

The three-magnon splitting (3MS) in SAF involves the annihilation of one of the optical magnons created by the antenna accompanied with the creation of a pair of magnons belonging to the acoustic branch, as sketched in Fig.~\ref{Fig_1_Setup_Spectrum_qualitative_3MS}(c) and (d). Such pair is detected as a \textit{doublet} of spectral lines at $\omega_\textrm{pump}/2 \pm \delta$ (blue lines) and at their harmonics, that adds to the contribution of the optical magnons that propagated without splitting [Fig.~\ref{Fig_1_Setup_Spectrum_qualitative_3MS}(d)].

\begin{figure} \hspace*{-0.5cm}  
\includegraphics[width=8.4 cm]{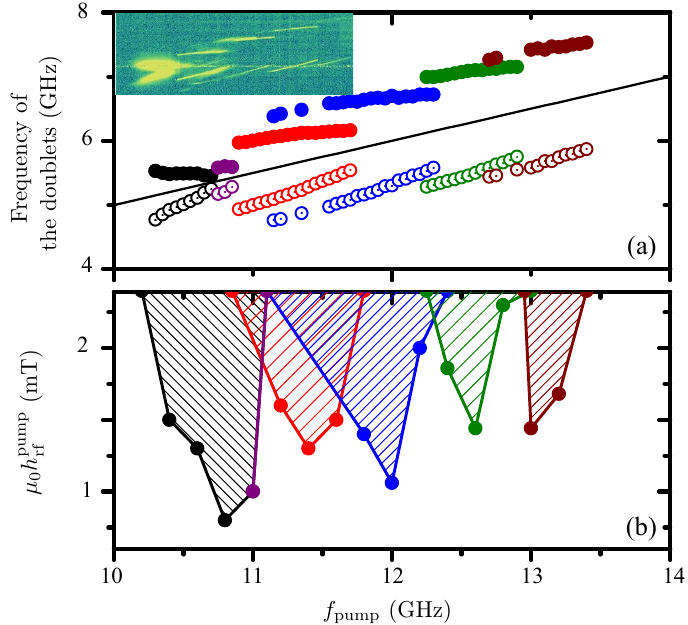}
\caption{Experimental results: (a) Frequencies $\omega_\textrm{ac,1}$ and $\omega_\textrm{ac,2}$ of the doublets of acoustic spin waves versus pump frequency for $\mu_0 h_x^\textrm{peak}=2.7$ mT. The modes are colored two-by-two to achieve $\omega_\textrm{ac,1}+\omega_\textrm{ac,2}=\omega_\textrm{pump}$.
A line is at $f_\textrm{pump}/2$ was superimposed for readability purposes. Inset: power spectral densities from which the frequencies are extracted (comparable interval of frequencies). (b) doublet-resolved 3MS thresholds forming the Arnold tongues.}
\label{Doublets_and_arnold_Tongues} \end{figure}

The frequencies of the created magnons vary with the pump frequency in a non-intuitive manner: this is illustrated in Fig.~\ref{Doublets_and_arnold_Tongues}(a) for a stripe of width $w_\textrm{stripe}=3~\mu$m. For some specific pump frequencies, several doublets can be observed at the same time [for instance at 11.5 GHz in Fig.~\ref{Doublets_and_arnold_Tongues}(a)]. When this happens, we pair the spectral lines in (colored) doublets such that the sum frequency of the doublet matches with $\omega_\textrm{pump}$, hence complying with energy conservation. \\

The 3MS processes occur when several conditions are met. The stimulus frequency $\omega_\textrm{pump}$ must be above the gap of the optical branch, i.e. above the frequency $\omega_\textrm{op0}$ of the uniform optical SW. Besides, the applied power must exceed a threshold which depends on $\omega_\textrm{pump}$. The dependence of the threshold with this frequency follows the classical shape of Arnold tongues seen in parametric pumping \cite{pikovsky_universal_2001} or injection locking \cite{martin_parametric_2011}, where each doublet can be assigned to its own Arnold tongue [Fig.~\ref{Fig_1_Setup_Spectrum_qualitative_3MS}(b)]. When scanning $\omega_\textrm{pump}$ the system enters and exits the successive (potentially overlapping) tongues: each doublet exists only in a restricted interval of pump frequencies. Increasing $\omega_\textrm{pump}$ thus leads to apparent jumps (discontinuities) of the frequency of the created acoustic magnons [inset in Fig.~\ref{Doublets_and_arnold_Tongues}(a)].

\begin{figure} 
\hspace*{-0.5cm}  \includegraphics[width=9.3 cm]{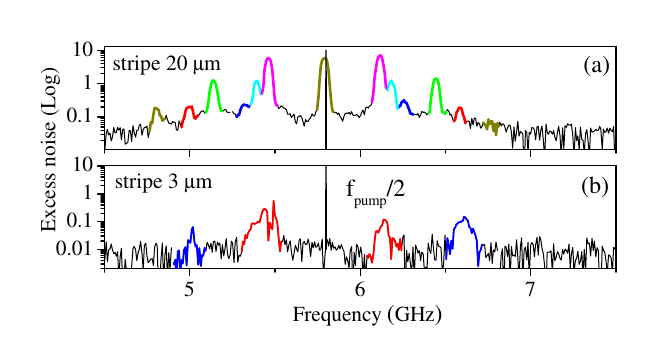}
\centering
\caption{Influence of stripe width on the frequencies of the acoustic doublets generated by 3MS for a pump at $2\times5.8$ GHz and a dc field of 33 mT. The peaks are colored by pairs according to energy conservation. (a) for a stripe width of 20 $\mu$m where 7 doublets are perceived, including a degenerate one. (b) for a stripe with of 3 $\mu$m where 2 doublets are perceived.}
\label{20microns_versus_3microns_spectra} \end{figure}

Because of the antenna geometry, the directly excited optical SWs are constrained to have a wavevector pointing in the $k_x$ direction. Since the dispersion relation of the optical mode $\omega_\textrm{op}(k_x)$ is continuous, the directly excited optical SWs can adapt to the pump frequency in a continuous manner. The jumps of the frequency of the created acoustic magnons are thus not likely to arise from the subsystem of the directly excited optical SWs.  
Conversely, the transverse part of the 3MS-generated acoustic SWs is constrained by stripe width and is therefore susceptible to quantization. To test if this is the cause for the jumps of the frequencies of the doublets [Fig.~\ref{Doublets_and_arnold_Tongues}(a)], we tuned the quantization condition by varying the stripe width. Fig.~\ref{20microns_versus_3microns_spectra}(a) compares the cases of $w_\textrm{stripe} = 3$ and 20 $\mu$m. While for the narrow stripe, at most two doublets could be observed at a time, up to 7 doublets can be observed for the largest stripe, with a much smaller frequency spacing. This argues also for a confinement-induced quantization of the split acoustic modes.

\begin{figure} \includegraphics[width=9 cm]{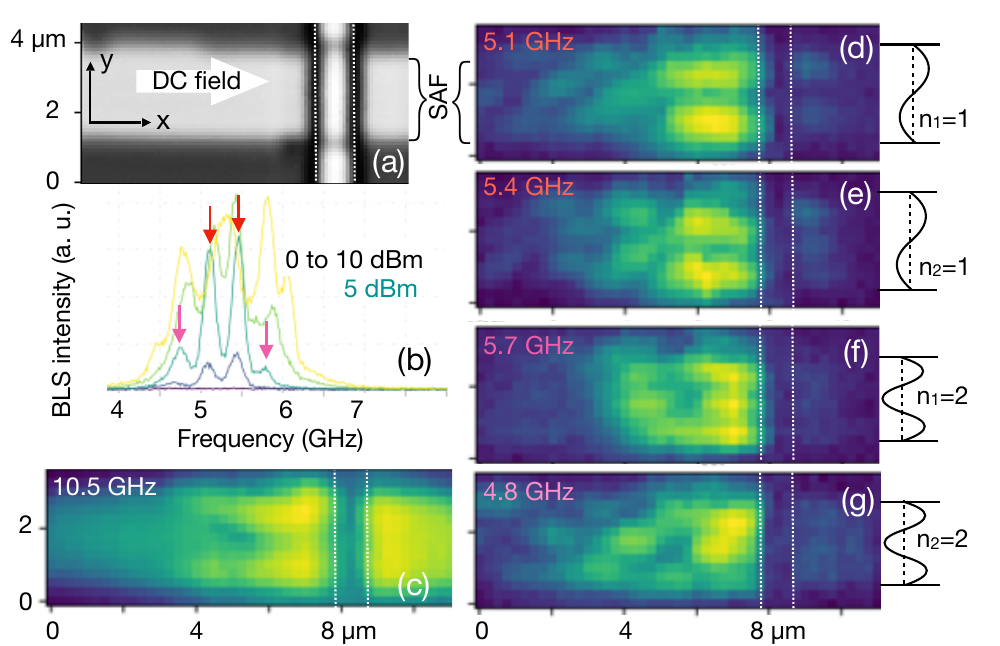}
\caption{Imaging of the modes involved in 3MS for $f_\textrm{stimulus}=10.5$ GHz, a field of 27 mT, a stripe width $\approx 2~\mu$m and an antenna (superimposed dotted lines) of width $\approx 1~\mu$m. (a): Optical image of the device. (b): $\mu$BLS spectra near $f_\textrm{stimulus}/2$ for powers $P_\textrm{source}=0$, 2.5, 5, 7.5 and 10 dBm. (c): $\mu$BLS image of the response at the applied frequency.
(d, e): Responses at 5.1 and 5.4 GHz  corresponding to $|\delta|=150~\textrm{MHz}$.
(f, g): Responses at 5.7 and 4.8 GHz ($|\delta|=450~\textrm{MHz}$).
Images (c-g) have a logarithmic color scale and are for $P_\textrm{source}=5~\textrm{dBm}$. Right sketches in (d-g): speculated wave profile in the transverse direction.   }
\label{BLSresults}
\end{figure}

To further confirm this speculation, we imaged the modes involved in 3MS using $\mu$BLS (Fig.~\ref{BLSresults}). This method uses a microscope to collect all the photons scattered by the SWs of a given frequency that are present at positions that are scanned across the sample. It therefore maps the local intensity of the SWs that share the same frequency. We report the experiment conducted on our narrowest stripe ($w_\textrm{stripe} \approx 2 ~\mu\textrm{m}$) because the frequency resolution of $\mu$BLS is insufficient to separate the doublets in wider stripes. At the stimulus frequency [Fig.~\ref{BLSresults}(b)] the directly excited spin waves have an optical character: they radiate energy towards \textit{both} sides of the antenna, as expected for modes having a tilted $\vee$-shaped dispersion relation \cite{millo_unidirectionality_2023}. The mode emitted towards $x>0$ is quasi-uniform within the width and has a long propagation length \footnote{Because of a parasitic coupling between the two antennas, the second antenna located at $x= 15\,\mu$m generates also a SW signal complicating the SW patterns. For didactic purpose, the images in Fig.~\ref{BLSresults}) are restricted to the vicinity of the first antenna.}. Conversely, two modes seem simultaneously emitted towards $x<0$: an intense uniform one also with a long propagation length, and a weaker unexpected one with a much shorter propagation length and two intensity lobes within the width. This weaker unexpected mode will be disregarded hereafter \footnote{One could think that this unexpected mode would be a parasitic second harmonic signature of the modes generated by 3MS. The profile observed in Fig.~\ref{BLSresults}(c) was found to persist below the threshold of 3MS, indicating that it is truly an optical mode directly emitted by the antenna. The direct excitation of this mode is not understood.}. 
The modes generated by 3MS are displayed in Figs.~\ref{BLSresults}(d, g). The acoustic character of the split modes is witnessed by their unidirectional propagation towards $x<0$. All observed split modes have a finite number $n \geq 1$ of nodes in the transverse direction. The transverse profile of the modes of the doublet nearest to $\omega_\textrm{pump}/2$ [Figs.~\ref{BLSresults}(d,e)] exhibits two lobes separated by one node (i.e. $n=1$). The next doublet involve acoustic modes with three lobes (i.e. $n=2$) of spacings close to the resolution of BLS [Figs.~\ref{BLSresults}(f, g)]. This proves that the generated acoustic modes have a quantized character within the stripe width.

To model the 3MS process we follow a rational based on the approximate 
dispersion relations of the acoustic mode found in ref.~\onlinecite{millo_unidirectionality_2023} and a quantization of the transverse profile (i.e. the $y$ dependence) of the acoustic SWs, as inferred from the BLS data.
When $\omega_\textrm{pump}= \omega_\textrm{op0}$, the antenna populates only one optical mode: the uniform ($\vec{k}_\textrm{opt} \approx \vec 0$) one that we shall sketch as $\lvert \vec 0 \rangle_\textrm{opt}$. Conversely, the antenna populates two initial states as soon as $\omega_\textrm{pump} > \omega_\textrm{op0}$. Their wavevectors are: 
\begin{equation} 
\vec  k_\textrm{opt} = \frac 1 {v_{g,x}^\textrm{op}}\big(\omega_\textrm{pump}-\omega_{\textrm{op}0}\big) \vec e_x \label{kopt}, \end{equation}
where $v_{g,x}^\textrm{op} \approx 13.4$ and -8.2 km/s are the group velocities of $k_x >0 $ or $k_x<0$optical branches (see Table I of ref.~\cite{millo_unidirectionality_2023}).
These states can be sketched as $\lvert\rightarrow\rangle_\textrm{opt}$ and $\lvert\longleftarrow\rangle_\textrm{opt}$, where the arrows illustrate the wavevectors in analogy to the plane waves in unbounded media.

The quantization of the acoustic mode relies on the analogy with the confined modes of a single-layer ferromagnetic thin stripe \cite{bayer_spin-wave_2006} with longitudinal magnetization. The confinement direction $(y)$ is the reciprocal direction of SAF SWs, allowing a priori for modes that have a standing wave character in this direction. 
We thus write an ad-hoc quantization ansatz that empirically accounts for the magnetization profile of the acoustic modes of the SAF. This ansatz postulates that the \textit{equivalent} transverse wavevector can be written as:
\begin{equation}
\vec k_{\textrm{ac,}\ell}.\vec e_y = \frac{n_\ell \pi }{{w}_\textrm{stripe}^\textrm{eff}} ,~~\textrm{with~}  n_\ell \in \mathbb Z  \label{quantization}
\end{equation}
where $| n_\ell |$ are the numbers of nodes of the transverse profiles of the two acoustic SWs $\ell=1, 2$ generated by 3MS.  $w_\textrm{stripe}^\textrm{eff} \approx w_\textrm{stripe}$ is a effective width that accounts for the dynamic dipolar pinning conditions at the stripe edges.

%

\begin{figure} \hspace*{-0.5cm}  \includegraphics[width=9.3 cm]{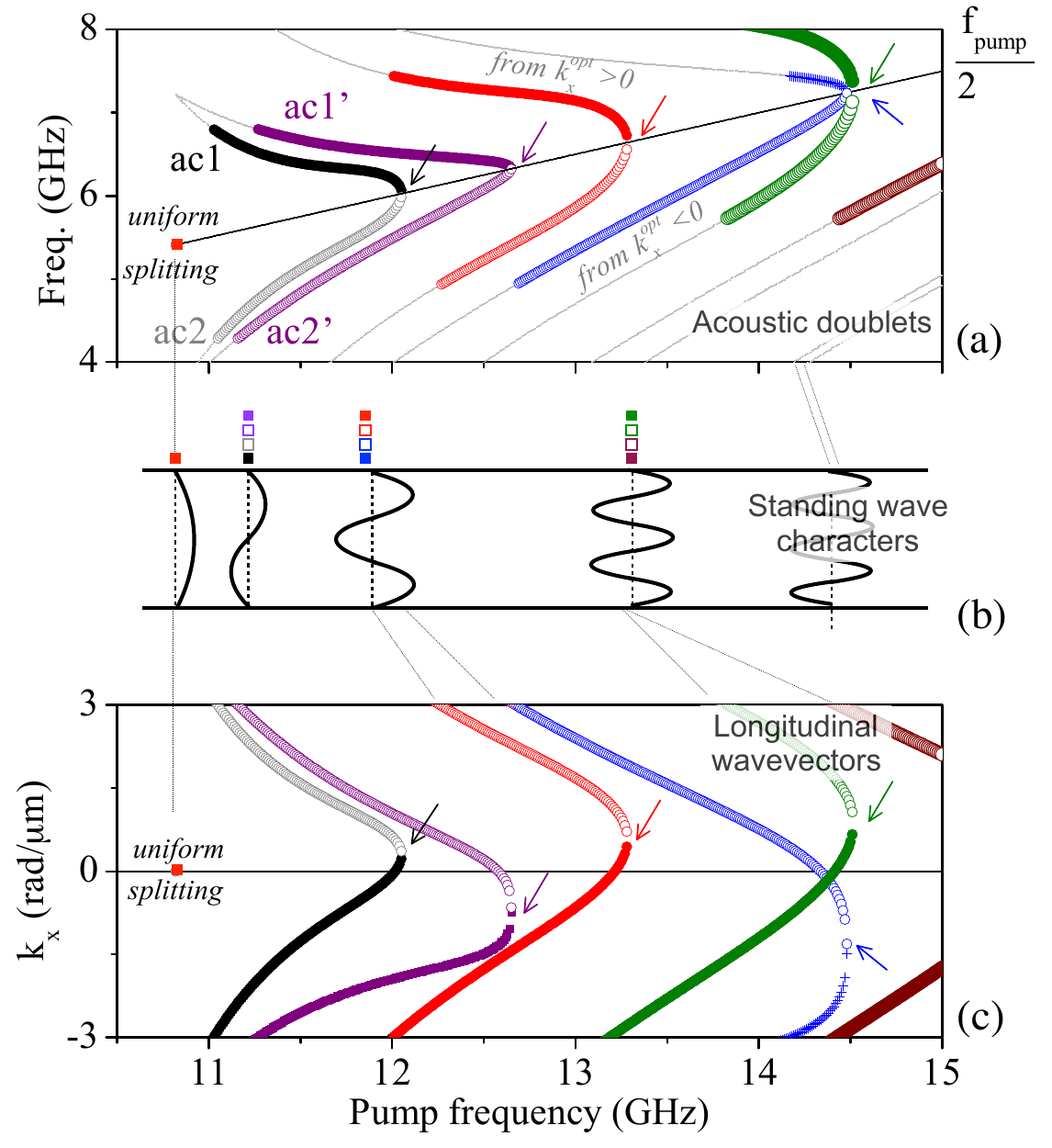}
\caption{Predicted 3MS channels in the conservation model for a 3 $\mu$m-wide stripe. The color code is that of the experiment of Fig. 2. 
The arrows denote the degenerate splittings. 
(a) Frequencies of the acoustic magnons resulting from splitting either  $\lvert\longleftarrow\rangle_\textrm{opt}$ [\textcolor{black}{blue and purple}, scenario of Fig.~1(c)] or $\lvert\rightarrow\rangle_\textrm{opt}$ [\textcolor{black}{black, green and wine}, scenario of Fig.~1(d)] optical magnons. A line at half of the pump frequency was superimposed for readability. The lines are bold when the SWs should be measurable by the antenna. (b) Sketches of the transverse profiles of the split magnons. (c) Longitudinal (i.e. propagative) component of the wavevectors $\vec k_\textrm{ac}.\vec e_x$  of the split magnons.}
\label{analytical_solutions} \end{figure}

Elucidating the 3MS mechanism is finding the pairs $\{\vec k_\textrm{ac,1} , \vec k_\textrm{ac,2}\}$ and their frequencies $\{\omega_\textrm{ac,1} , \omega_\textrm{ac,2}\}$. Since the system is time-invariant, the pair must satisfy energy conservation:
\begin{equation}
\omega_\textrm{ac,1}+ \omega_\textrm{ac,2} =  \omega_\textrm{pump}.\label{energyConservation} \end{equation}
Besides, we will assume that the total momentum must also be conserved, even for the effective transverse part of the wavevectors despite the absence of translational invariance in this direction: 
\begin{equation} \vec  k_\textrm{opt} =  \vec k_\textrm{ac,1} + \vec k_\textrm{ac,2}. \label {MomentumConservation} \end{equation}
Eqs.~\ref{kopt} and \ref{MomentumConservation} entail $n_1=-n_2$, in line with the experimental findings.  Note that the created acoustic magnons may have opposite $k_x$ components. They however both radiate energy ("propagate") towards the $x<0$ side thanks to the unidirectionality of these acoustic SWs, and can therefore be measured by the same receiving antenna as seen experimentally. For each $\omega_\textrm{pump}$ and each assumed $n_1$ in Eq.~\ref{quantization}, we are left with two equations (Eq.~\ref{energyConservation} and the $x$ components of Eq.~\ref{MomentumConservation}) for two remaining unknowns (the $x$ components of $\vec k_\textrm{ac,1}$ and $\vec k_\textrm{ac,2}$), which yields a unique solution. 
Such solutions are displayed in Fig.~\ref{analytical_solutions} for a stripe width of $w_\textrm{stripe}=3~\mu$m. The 3MS scenarios appear to depend substantially on the pump frequency.

When $\omega_\textrm{pump}= \omega_\textrm{op0}$, the initial optical mode $\lvert \vec 0 \rangle_\textrm{opt}$ is allowed to split into two identical acoustic modes that are uniform: $\lvert \vec 0 \rangle_\textrm{ac}$ and $\lvert \vec 0 \rangle_\textrm{ac}$. This \textit{"uniform"} splitting is limited to this single stimulus frequency $\omega_\textrm{pump}= \omega_\textrm{op0}$. Alternatively, $\lvert \vec 0 \rangle_\textrm{opt}$ can split into two non-degenerate acoustic modes that have opposite $k_x$'s and nodes in the transverse direction (i.e. $|n_1|=|n_2|\neq 0$). The frequency splitting $\omega_\textrm{ac,1}-\omega_\textrm{ac,2}$ is then related to these opposite $k_x$'s and the non-reciprocity in this direction. This non-degenerate splitting from $\lvert \vec 0 \rangle_\textrm{opt}$ is predicted to yield acoustic SWs with wavevectors $k_x$ that are too large to be measurable by the antenna, already for $n=1$.  

When increasing the pump frequency to $\omega_\textrm{pump} > \omega_\textrm{op0}$, the stimulus now populates the modes $\lvert\longleftarrow\rangle_\textrm{opt}$ and $\lvert\rightarrow\rangle_\textrm{opt}$ that are both susceptible to splitting. 
All resulting split modes are predicted to have a standing wave character in the stripe width and a plane wave character in the length. The splitting is in general non-degenerate: $
\vec k_\textrm{ac,1}.\vec e_x \neq \vec k_\textrm{ac,2}.\vec e_x$ and $\omega_\textrm{ac,1} \neq\omega_\textrm{ac,2}$. 
Keeping the analogy with plane waves, the 3MS can most often be sketched as follows: 
\begin{subequations}
\begin{align} 
 & \lvert\longleftarrow\rangle_\textrm{opt}  \textrm{~splits~}  \textrm{to} ~\lvert\nearrow\rangle + \lvert\searrow\rangle \textrm{~and~} \lvert\nwarrow\rangle + \lvert\swarrow \rangle \label{kopnneg}\\
 & \lvert\rightarrow\rangle_\textrm{opt} \textrm{~splits~}  \textrm{to~another}~\lvert\nwarrow\rangle + \lvert\swarrow\rangle \textrm{~and~} \lvert\nearrow\rangle + \lvert\searrow \rangle
\end{align}
\end{subequations} 
where the first (second) channels are plotted as purple (black/gray) doublets in Fig.~\ref{analytical_solutions}. Each doublet forms a kind of rotated chevron pattern (i.e. "$>$") in frequency versus frequency space [Fig.~\ref{analytical_solutions}(a)]. The split magnons may either be observable experimentally (bold symbols in Fig.~\ref{analytical_solutions}) or fall out of the wavevector detection range of the experiment (narrow lines). For $w_\textrm{stripe}=3~\mu$m the conservation model (Eqs.~\ref{energyConservation}-\ref{MomentumConservation}) and the experimental detection limit ($|k_x| \leq k_\textrm{max}^\textrm{antenna}$) predicts the possibility to observe at most two doublets simultaneously, as in experiments. For wider stripes, more states are available to scatter into and the number of observable doublets is predicted to increase accordingly, leading to a dense pattern of chevrons (not shown). A degenerate splitting is allowed only at some very specific stimulus frequencies (arrows in Fig.~\ref{analytical_solutions}). In these cases the split modes are copropagating waves that can be sketched as: $\lvert\textrm{ac1}\rangle=\lvert\textrm{ac2}\rangle=\lvert\nwarrow\rangle + \lvert\swarrow\rangle$ if $k_x^\textrm{opt} <0$ or as $\lvert\textrm{ac1'}\rangle=\lvert\textrm{ac2'}\rangle=\lvert\nearrow\rangle + \lvert\searrow\rangle$ if $k_x^\textrm{opt} >0$.

Several outcomes of the conservation model are worth noticing. Upon increasing $\omega_\textrm{pump}$, the extra energy put in the optical mode requires ever larger wavevectors $|k_{x}^\textrm{op}|$. Since the longitudinal group velocities obey $|v_{g, x}^\textrm{op}| > |v_{g, x}^\textrm{ac}|$, acoustic modes with ever increasing number of nodes $|n_{1,2}|$ in the transverse ($y$) direction are needed to satisfy the energy conservation that cannot be satisfied by the $k_x$ components alone. Because of the quantized nature of the $n$'s, the split acoustic modes therefore jump from one quantized mode to the next one upwards in the energy scale as $\omega_\textrm{pump}$ increases. The jumps are expected to progressively shrink to zero when enlarging the stripe and lifting the confinement contraint.
The doublets are predicted to form a series of separated chevrons (i.e. $\gg$-like) in agreement with the experiment. 
The frequencies of the high-frequency and low-frequency members of each doublets are predicted to converge to degenerate splitting when increasing $\omega_\textrm{pump}$, until this 3MS channel cease to exist because the conservation laws can no longer be satisfied. This bears some similarity with the experimental behavior, although in experiment on narrow stripes the 3MS channel deactivates before the convergence to degenerate splitting (Fig.~\ref{Doublets_and_arnold_Tongues}). 

Some of the splitting channels allowed by the conservation model (Eqs~\ref{energyConservation}- \ref{MomentumConservation}) are not observed experimentally. Notably, the splitting of the uniform optical mode into two uniform acoustic mode was not observed. This is unlikely to be a question of sensitivity or insufficient rf field. Rather, we believe that since the phase space where this splitting can be observed is a single point, it disappears as soon as the condition $\omega_\textrm{op0}=2 \omega_\textrm{ac0}$ is not strictly met, for instance because of a tiny error in the applied field. \\
Another deficiency of the model is that it predicts 6 observable doublets [Fig.~\ref{analytical_solutions}(a)] while only 5 are measured [Fig.~\ref{Doublets_and_arnold_Tongues}(a)] for the 3 $\mu$m stripe. Note that the first two chevrons are predicted to be very close in frequency. In experiments the modes have linewidths of typically 300 MHz so that two such predicted chevrons may be broadened accordingly, up to a point where they cannot be resolved. We believe that this is the likely reason why experimentally, the first chevron appears with a much wider width than any other one [see inset in Fig.~\ref{Doublets_and_arnold_Tongues}(a)]. \\
Finally, the model allows the splitting into degenerate pairs (i.e. $\omega_\textrm{ac,1}=\omega_\textrm{ac,2}$) when $\omega_\textrm{pump} > \omega_\textrm{op0}$. This was observed experimentally only in the widest stripes [Fig.~\ref{20microns_versus_3microns_spectra}(a)]. It is possible that this degenerate splitting has a larger threshold in narrower stripes and appears only at powers inaccessible with our present instrumentation.
As a last remark, we confess that our measurements cannot exclude the possible presence of the subsequent scattering processes. The satisfactory agreement between experiment and theory however supports the interpretation that the observed multiple pairs primarily originate from the geometric confinement that quantizes the acoustic modes within the stripe. As a perspective, our method opens the possibility to study the phase coherence of magnons created by 3MS, a study rendered much easier by the fact that created magnons propagate in the same direction.

Our findings have implications for the applications on non-linear spin waves, because SW frequencies can undergo transformations in ways that are uncommon in microwave technology. Since the input and output frequencies of our device are typically incommensurate, we anticipate applications in microwave frequency manipulation, in particular for performing frequency conversion without the need of a microwave mixer and a local oscillator. The concept could be extended to much higher frequencies by using easy-plane antiferromagnets like hematite \cite{lebrun_long-distance_2020, schonfeld_dynamical_2025} at the cost of Tesla-scale applied fields. 

\begin{acknowledgments}
This work was supported by the French RENATECH network, by the French National Research Agency (ANR) as part of the “Investissements d’Avenir” and France 2030 programs. This includes the MAXSAW project ANR-20-CE24-0025 and the PEPR SPIN projects ANR 22 EXSP 0008 and ANR 22 EXSP 0004.  We also acknowledge support from the EU Research and Innovation Programme Horizon Europe under grant agreements n°101070290 (NIMFEIA). We thank S. M. Ngom for sample fabrication.
\end{acknowledgments}

\section*{DATA AVAILABILITY} 
The data that support the findings of this article are openly available at \cite{devolder_datasets_2025}.

\end{document}